\documentclass[conference]{IEEEtran}
\IEEEoverridecommandlockouts

\usepackage[boxruled]{algorithm2e}
\usepackage{enumerate}
\usepackage{multirow,array}
\usepackage{dirtytalk}
\usepackage{cite}
\usepackage{graphicx}
\usepackage{psfrag}
\usepackage{caption}
\usepackage{subcaption}
\usepackage{url}
\usepackage{amsmath}
\usepackage{amsthm}
\usepackage{kbordermatrix}
\usepackage{array}
\usepackage{algorithm2e}
\usepackage{amssymb}
\usepackage{amsfonts}
\usepackage{float}
\usepackage{textcomp}
\usepackage{xcolor}
\usepackage{tabu}
\usepackage{array}
\newcolumntype{P}[1]{>{\centering\arraybackslash}p{#1}}
\newcolumntype{M}[1]{>{\centering\arraybackslash}m{#1}}

\usepackage{dsfont}

\usepackage{amsmath}
\usepackage{xparse}

\ExplSyntaxOn
\NewDocumentCommand{\RN}{m}
 {
  \textup{ \int_to_Roman:n { #1 } }
 }
\ExplSyntaxOff

\def\BibTeX{{\rm B\kern-.05em{\sc i\kern-.025em b}\kern-.08em
    T\kern-.1667em\lower.7ex\hbox{E}\kern-.125emX}}
\usepackage{float}

\renewcommand{\boxed}[1]{\text{\fboxsep=.2em\fbox{\m@th$\displaystyle#1$}}}

\newtheorem*{conjecture*}{Conjecture}
\newtheorem{lemma}{Lemma}

\newtheorem{corollary}{Corollary}
\newtheorem{remark}{Remark}
\newtheorem{definition}{Definition}
\newtheorem{theorem}{Theorem}
\newtheorem{example}{Example}

\title{Low Complexity Distributed Computing via Binary Matrices with Extension to Stragglers}
\begin{document}

\author{
  \IEEEauthorblockN{Shailja Agrawal, Prasad Krishnan}
  \IEEEauthorblockA{
                    International Institute of Information Technology, Hyderabad\\ 
                    Email: \{shailja.agrawal@research. , prasad.krishnan@\}iiit.ac.in}
\vspace{-0.2cm}
}

\maketitle
\begin{abstract}
We consider the distributed computing framework of MapReduce, which consists of three phases, the Map phase, the Shuffle phase and the Reduce phase. For this framework, we propose the use of binary matrices (with $0,1$ entries) called \textit{computing matrices} to describe the map phase and the shuffle phase. Similar binary matrices were recently proposed for the coded caching framework. The structure of ones and zeroes in the binary computing matrix captures the map phase of the MapReduce framework. We present a new simple coded data shuffling scheme for this binary matrix model, based on a \textit{identity submatrix cover} of the computing matrix. This new coded shuffling scheme has in general a larger communication load than existing schemes, but has the advantage of less complexity overhead than the well-known earlier schemes in literature in terms of the file-splitting and associated indexing and coordination required. We also show that there exists a binary matrix based distributed computing scheme with our new data-shuffling scheme which has strictly less than twice than the communication load of the known optimal scheme in literature. The structure of this new scheme enables it to be applied to the framework of MapReduce with stragglers also, in a straightforward manner, borrowing its advantages and disadvantages from the no-straggler situation. Finally, using binary matrices derived from combinatorial designs, we show specific classes of computing schemes with very low \textit{file complexity} (number of subfiles in the file), with marginally higher communication load compared to the optimal scheme for equivalent parameters.

\end{abstract}

\section{Introduction}

One of the most popular distributed computing framework is MapReduce \cite{mapreduce}. MapReduce is a programming model and enables the processing of large data sets on distributed servers. In the MapReduce framework, the large data file is partitioned into smaller parts, and these parts are then assigned to different servers for processing in a distributed fashion. There are two main phases in MapReduce : \textit{map} and \textit{reduce}, and a third \textit{data shuffling} phase connects the two. In the map phase, each of the data parts is processed by one or more servers to generate some intermediate values (IVAs) using map functions. In the next step, servers exchange these IVAs (called \textit{data shuffling}) so that the final outputs can be calculated in a distributed fashion across the server using the {reduce functions}.

As observed in \cite{33per}\cite{70per}, data shuffling is a significant phase in determining the performance of the original MapReduce framework, which passes the IVAs in an uncoded manner during the shuffling phase.  In \cite{CMR}, it was shown that it is possible to code the IVAs together before the shuffling process by exploiting the fact that $r$ distinct carefully chosen nodes are mapping the same subfiles (and hence have the same IVAs). This leads to great savings in the communications load. The parameter $r$ is known as the \textit{computation load}, and indicates that the price to pay for reducing the communications load. This new framework with a coded shuffling phase, is known as Coded MapReduce. This model was further studied in \cite{RDC}, where it was shown that the communication load achieved by the Coded MapReduce scheme of \cite{RDC} is optimal.  
In \cite{CommunicationComputationAlternativetradeoff,StorageComputationISWCS,QYDC}, the model proposed in \cite{CMR,RDC} was further extended to consider coded MapReduce schemes in which the nodes need not compute IVAs of all the stored subfiles for completing their reduce tasks. Tradeoffs between storage, computation, and communication loads were derived in \cite{StorageComputationISWCS,QYDC}, and an optimal scheme which meets this tradeoff was also presented. 

Coded distributed computing schemes in the presence of \textit{stragglers} in the computation process were studied in \cite{Speedingup,UnifiedCodingFramework} for the case of computing functions which are linear. Stragglers are nodes which are either slow or completely unable to complete their map tasks. Subsequently, the works \cite{QYS,vinayakPDA}, extended the coded MapReduce model of \cite{RDC} to arbitrary function computation  in the presence of full and partial stragglers. 

The contributions and organization of this work are as follows. After a brief review of the coded MapReduce setup in Section \ref{systemmodel}, we introduce the notion of binary computing matrices for distributed computing in Section \ref{binarymatricesforcomputing}. For this setup, we propose a new simple delivery scheme in Section \ref{lowcomplexitydatashufflingsubsection}. We interpret the optimal coded MapReduce scheme shown in \cite{RDC} as a binary matrix based scheme, and show that the load achieved by our data-shuffling scheme for the same is strictly less than twice that of the optimal load. However, our scheme has lesser complexity in the data shuffling phase, as it avoids the splitting the IVAs further into smaller packets. We discuss these in Section \ref{advantagessubsubsection}. We also present binary computing matrix constructions from several combinatorial designs and show their parameters, including the load obtained using our scheme. Compared to the optimal scheme, these schemes give a larger communication load,  but have very low file complexity (Section \ref{combinatorialdesigns}). By default, our new scheme does not ensure \textit{communication load balancing}, i.e., not all servers participate in the transmissions during the data-shuffling phase; but such load balancing can be achieved by finding perfect matchings on an appropriately defined graph (Section \ref{loadbalancingsubsec}). Interestingly, the load imbalanced feature of our raw scheme can be exploited in order to protect against stragglers. In Section \ref{fullstragglersec}, we extend our scheme to distributed computing with stragglers. We assume that some nodes (numbering upto some threshold based on the properties of the binary computing matrix) are unavailable or failed. For this full straggler model, the communication load in our scheme is increased by a factor that depends only on the increase in the number of functions reduced per node. We present numerical comparison of our straggler-robust scheme with the optimal scheme for this setting from \cite{QYS}. 

\textit{Notations and Terminologies: } For any positive integers $l,m$ s.t $m\geq l$, we use notation $[m] \triangleq \{1,\hdots,m\}$ and $[l:m]\triangleq \{l,l+1,\hdots,m\}$. For a matrix $A$ whose rows are indexed by a finite set ${\cal R}$ and columns are indexed by a finite set ${\cal C}$, the element in the $r^{th}$ row  $(r \in {\cal R})$ and $l^{th}$ column  $(l \in {\cal C})$ is denoted as $A(r,l)$. Binomial coefficient is denoted by $\binom{n}{r} \triangleq \frac{n!}{r!(n-r)!}$ for $n\geq r \geq 0$. $P\backslash Q$ denotes the elements in $P$ but not in $Q$, where $P$ and $Q$ are sets. For some element $i$, we also denote $P\backslash \{i\}$ by $P\backslash i$. $a\oplus b$ represents the bit-wise xor operation between two binary vector $a$ and $b$ of same length.

\section{System Model}
\label{systemmodel}
We briefly review the distributed computing framework presented in \cite{RDC}, where the task is to compute $Q$ output functions on a large file using $K$ servers or computing nodes, which are indexed by a set ${\cal K}$. As in \cite{RDC}, we assume that $Q\geq K $ and $\frac{Q}{K}$ is an integer. The file is divided into $N$ subfiles (we assume $N\geq K$), and denoted by a set ${\cal N}$ of size $N$. The parameter $N$ is also referred as \textit{file complexity} in this work. Each subfile is assigned to $r$ servers where $r$ is called \textit{computational load}. Clearly, $r\geq 1$. Let ${\cal M}_k \subseteq {\cal N}$ denote the set of the subfiles assigned to server $k$, $k\in {\cal K}$.

Let the $Q$ output functions be denoted as $\phi_1,\hdots,\phi_Q$ where each $\phi_q: q \in [Q]$ maps all the input files to $u_q$, where $u_q = \phi_q(\{f\in{\cal N}\})$ is a binary stream of some fixed length. Let $g_{q,f} , ~q\in [Q], f\in {\cal N}$ denote the \textit{map functions} which maps the input subfile $f\in {\cal N}$ into $Q$ intermediate values (IVAs), each a consisting of $T$ bits, denoted as $\{v_{1,f},v_{2,f}.\hdots,v_{Q,f} \}$. Each $v_{q,f}\triangleq  g_{q,f}(f), ~q\in [Q], f\in {\cal N} $ represents the IVA of length $T$ bits of the corresponding to the $q^{th}$ function and the subfile $f$. The \textit{reduce function} denoted by $h_q , q\in[Q]$ maps IVAs $v_{q,f}: f\in {\cal N}$ to output bit stream $u_q$. Thus we have $u_q = \phi_{q}(\{f\in {\cal N}\}) = h_q(\{v_{q,f}:f\in {\cal N}\}) = h_q(\{g_{q,f}(f):f\in {\cal N}\})$.

A distributed computing scheme consists of three phases : map, shuffle and reduce phases, which we describe using the above functions as follows. 

\subsection{Map Phase} In map phase, each server $k\in {\cal K}$ will compute all the IVAs for the subfiles in ${\cal M}_k$ using the map functions, i.e., server $k$ computes $g_{q,f}(f):f\in{\cal M}_k,\forall q$. Thus after map phase, server $k\in {\cal K}$ has $\{v_{q,f} : q\in [Q], f\in {\cal M}_k \}$. 

\subsection{Shuffle Phase} Each server is responsible for reducing $\beta\triangleq \frac{Q}{K}$ functions. Let ${\cal W}_{k}\subset [Q]$ denote the indices of the functions evaluated (reduced) at server $k\in {\cal K}$, where $\cup_{k=1}^K{\cal W}_k=[Q]$ . For a server to compute the output of a reduce function, it needs the IVAs of that output function for all the subfiles. Thus each server $k\in {\cal K}$ requires $\{v_{q,f} : q\in {\cal W}_{k} , f\not\in {\cal M}_k \}$ to reduce the functions assigned to it. Hence, in the shuffle phase, the servers send broadcast transmissions to each other to make sure that each server receives the IVAs it needs for performing the reduce operations assigned to it. 

\subsection{Reduce Phase} 
With the received IVAs in the shuffle phase and the IVAs computed locally in the map phase, server $k$ uses the reduce functions $h_q$ to compute the task assigned to it, i.e., the node $k$ computes $h_q(\{v_{q,f}: f\in {\cal N}\})$ for each $q\in{\cal W}_k$.

As in \cite{RDC} the normalized communication load $L$, of a distributed computing framework is defined as the (normalized) total number of bits communicated in shuffle phase by all the $K$ servers and can be calculated as 
\[
\text{}~L = \small \frac{\text{Total number of bits transmitted in shuffle phase}}{\text{$QNT$}},
\]
where $T$ is the size of each IVA in bits. The coded distributed computing framework introduced in \cite{CMR}, used coded transmission in the shuffle phase to reduce the communication load by doing extra computations. The communication load achieved by the scheme in \cite{CMR,RDC} for computational load $r$ is shown to be $\frac{1}{r}(1-\frac{r}{K})$ and this communication load is shown to be optimal in \cite{RDC}.

\section{Binary matrices and Distributed Computing}
\label{binarymatricesforcomputing}
In \cite{SSP} we used binary matrices to design coded caching schemes. In a similar vein, in this section we describe how a distributed computing scheme can be derived from a binary matrix with constant column weight. 
\begin{definition}
[Binary Computing Matrix] Consider a matrix $C$ with entries from $\{0,1\}$ with rows indexed by a $K$-sized set ${\cal K}$ and columns indexed by a $N$-sized set ${\cal N}$  such that the number of $0$'s in any column is constant (say $r$).
Then the matrix $C$ defines a distributed computing scheme with $K$ users (indexed by ${\cal K}$), file complexity $N$ (subfiles indexed by ${\cal N}$) and computation load $= r$ as follows:
\begin{itemize}
\item Server $k\in {\cal K}$ maps subfile $f: f \in {\cal N}$ if $C(k,f) = 0$ and does not map it if $C(k,f) = 1$.
\end{itemize}
We then call the matrix $C$ as a $(K,N,r)$-binary computing matrix. 
\end{definition}

In \cite{SSP}, combinatorial designs were used to construct binary matrices which were used in coded caching. We discuss one example of such a matrix in this section, and use it to also describe our new coded shuffling scheme. Later, in Section \ref{combinatorialdesigns}, we present several constructions of binary computing matrices derived from combinatorial designs, with their associated distributed computing parameters. 
\begin{example}
\label{DC_matrix}
Consider a set system $({\cal K},{\cal N})$ given by 
\newline ${\cal K} = \left\{1,2,3,4,5,6,7 \right\}$
\newline ${\cal N} = \{127,145,136,467,256,357,234 \},$ where each element of ${\cal N}$ indicates a subset of ${\cal X}$ (for instance $127$ stands for $\{1,2,7\}$). 

Consider a matrix $C$ of size with rows indexed by ${\cal K}$ and columns indexed by ${\cal N}$, such that for each $k\in{\cal K},f\in{\cal N}$, $C(k,f)=1$ if and only if $k\in f$. Thus, $C$ is an incidence matrix for this set system, and is written as
\renewcommand{\kbldelim}{(}
\renewcommand{\kbrdelim}{)}
\[
  \text{$C$} = \kbordermatrix{
    & 127 & 145 & 136 & 467 & 256 & 357 & 234 \\
    1 & 1 & 1 & 1 & 0 & 0 & 0 & 0\\
    2 & 1 & 0 & 0 & 0 & 1 & 0 & 1\\
    3 & 0 & 0 & 1 & 0 & 0 & 1 & 1\\
    4 & 0 & 1 & 0 & 1 & 0 & 0 & 1\\
    5 & 0 & 1 & 0 & 0 & 1 & 1 & 0\\
    6 & 0 & 0 & 1 & 1 & 1 & 0 & 0\\
    7 & 1 & 0 & 0 & 1 & 0 & 1 & 0
  }
\]
It is easy to see that gives us a $(7,7,4)$-binary computing matrix. The corresponding distributed  computing system has $K=7$ nodes, $N=7$ subfiles, in which each subfile is stored in $r=4$ users. For instance, subfile indexed by $127$ is stored in users $3,4,5$ and $6$. Incidentally, this corresponds to a combinatorial design known as $(7,3,1)-$ balanced incomplete block design (BIBD). Other combinatorial designs lead to such binary matrices as well (for example, see \cite{SSP}). 
\end{example}


In map phase, each server $k\in {\cal K}$ will compute the IVAs of all functions for all the subfiles for which entry $C(k,f)=0$ in computing matrix. At the end of the map phase, server $k\in{\cal K}$ already has $\{v_{q,f} : C(k,f)=0 \}$ and it needs $\{v_{q,f} : C(k,f)=1 \}$ for each $q\in{\cal W}_k$ from the shuffle phase. 

Let ${\cal W}_k = \{q_{(k,b)}~: \forall b\in [\beta] \}$ denote the functions to be reduced by server $k$.
In order to describe the shuffle phase in which we do coded transmissions, we first describe a single round of two transmissions based on the above described matrix based distributed computing scheme, which will serve a number of servers. 
Note that a submatrix of $C$ can be specified by a subset of the row indices ${\cal K}$ and a subset of column indices ${\cal N}$. We now recall the idea of an \textit{identity submatrix} of matrix $C$.

\begin{definition}\cite{SSP}
[Identity Submatrix] An $l \times l$ submatrix $C'$ of the matrix $C$ is an identity submatrix of size $l$ if its columns correspond to the identity matrix of size $l$ permuted in some way.
\end{definition}

The following lemma describes one transmission round corresponding to a given identity submatrix, consisting of two transmissions which exchanges the IVAs among the users indexing the rows of the submatrix.

\begin{lemma}
\label{lemma_identitysubmatrix}
Consider an identity submatrix of $C$ given by rows $\{k_1,k_2,..,k_l: k_i \in {\cal K}\}$ and columns $\{f_1,f_2,..,f_l: f_i \in {\cal N}\}$, such that $C(k_i,f_i)=1, \forall i \in [l]$, while $C(k_i,f_j)=0, \forall i,j \in [l]$ where $i\neq j$. Then there exists two transmissions of length $\beta T= \frac{QT}{K}$ bits each, one coded and one uncoded,  done by any two different servers $k_{i}$ and $k_{j} : i,j\in [l], i\neq j$ such that each server $k_{i}: i\in [l]$  can recover the missing IVAs 
$${ v_{q_{(k_{i},b)},f_{i}} : q_{(k_{i},b)}\in {\cal W}_{k_i}, }$$ 
for each $~f_{{i}}\not\in {\cal M}_{k_i},~\forall b\in[\beta]$, from these two coded transmissions.
\end{lemma}
\begin{IEEEproof}
By definition of identity submatrix, for each $i \in [l]$ the IVAs  $v_{q_{(k_{i},b)},f_{i}} : q_{(k_{i},b)}\in {\cal W}_{k_i}, ~f_{{i}}\not\in {\cal M}_{k_i} $ are not available at server $k_i$ but are available at the other servers $\{k_1,k_2,\hdots,k_l\} \backslash k_i$.
Therefore, for some $p\in [l]$, consider the coded transmission of length $\beta T$ by the server $k_p$
\[
\left\{
\sum\limits_{\substack{i=1 \\ i\neq p}}^l {v_{q_{(k_{i},1)},f_{i}}},
\sum\limits_{\substack{i=1 \\ i\neq p}}^l {v_{q_{(k_{i},2)},f_{i}}}, \hdots,
\sum\limits_{\substack{i=1 \\ i\neq p}}^l {v_{q_{(k_{i},\beta)},f_{i}}}.
\right\}
\]
From the above transmission,  each $i \in [l] \backslash p$ can clearly recover the IVAs $v_{q_{(k_{i},b)},f_{i}} : q_{(k_{i},b)}\in {\cal W}_{k_i}, f_{{i}}\not\in {\cal M}_{k_i} $. For instance, from transmission $\sum\limits_{\substack{i=1 \\ i\neq p}}^l {v_{q_{(k_{i},1)},f_{i}}}$, server $k_j: j\in [l]\backslash p$ can recover the intermediate value $v_{q_{(k_{j},1)},f_{j}}$ as all the other intermediate values ${v_{q_{(k_{i},1)},f_{i}}}: i\in [l]\backslash \{j,p\}$ are already present at server $k_j$ from map phase. Now, pick any server $k_i : i \in [l], i \neq p$. Let this server $k_i$ transmit the uncoded transmission of size $\beta T$ bits as follows.
\[
\{ v_{q_{(k_{p},1)},f_{p}} ,~ v_{q_{(k_{p},2)},f_{p}}, \hdots, v_{q_{(k_{p},\beta)},f_{p}} \}.
\]
Thus, server $k_p$ receives the IVAs $v_{q_{(k_{p},b)},f_{p}} : q_{(k_{p},b)}\in {\cal W}_{k_p}$ uncoded. This completes the proof.
\end{IEEEproof}

We shall use Lemma \ref{lemma_identitysubmatrix} to describe the complete shuffle phase. For that purpose we introduce few more terminologies, mostly borrowed from \cite{SSP}. 

For a computing matrix $C$, suppose $C(k,f) =1$ for some $k \in {\cal K}$ and $f \in {\cal N}$. The entry $C(k,f) =1$ is said to be \textit {covered} by the identity submatrix $B$ if $k$ and $f$ correspond to some row and column index of $B$ respectively.
\begin{definition}\cite{SSP}[Identity Submatrix Cover]
Consider a set $\mathfrak {C}=\{C_1,...,C_S\}$ consisting of $S$ identity submatrices of a computing matrix $C$ such that any $C(k,f) =1$ in $C$ is \textit {covered} by atleast one $C_i\in{\mathfrak{C}}$ . Then, $\mathfrak {C}$ is called an Identity Submatrix Cover of $C$.

\end{definition}

We also need the idea of an \textit{overlap}\cite{SSP} between identity submatrices of $C$, which enables us to calculate the communication load of the coded shuffling phase.
\begin{definition}
[Overlap] An \underline{overlap} in a collection of given identity submatrices is said to occur when some entry $C(k,f)=1$ in matrix $C$ is covered by more than one identity submatrix in the collection. If there are no overlaps in collection, we call it a collection of non-overlapping identity matrices.
\end{definition}
\subsection{A new simple low complexity coded data shuffling algorithm}
\label{lowcomplexitydatashufflingsubsection}
We are now ready to describe our new simple algorithm for the coded shuffling phase.
The following theorem captures this result.

\begin{theorem}
\label{DC_maintheorem}
Consider a binary computing matrix $C$ of size $K\times N$ with a non-overlapping identity submatrix cover $\mathfrak{C}= \{C_1,C_2,..,C_S\}$ where the size of each identity submatrix is $g\geq 2$. Then, there exists a distributed computing scheme with $K$ nodes, attaining computation load $r$ and communication load $L={\frac{2}{g}}{\left( 1- \frac{r}{K}\right)}$, with file complexity $N$.
\end{theorem}
\begin{IEEEproof}
By Lemma \ref{lemma_identitysubmatrix}, corresponding to each identity submatrix $C_i$ in $\mathfrak{C}$, there are two transmissions which exchanges all the missing IVAs with respect to the subfiles corresponding to columns of the submatrix $C_i$ amongst the users indexed by the rows of $C_i$. Note that since $\mathfrak{C}$ is an identity submatrix cover, each missing IVA at any user will be part of some transmission corresponding to some identity submatrix in $\mathfrak{C}$. Thus, all the missing IVAs at all the users corresponding to all the functions to be reduced, are decoded. Therefore the reduce functions can also be successfully executed at the respective nodes. As $\mathfrak{C}$ contains $S$ identity submatrices, the total number of transmissions are $2S$.  Each transmission is of size $\beta T=QT/K$ bits. Hence the communication load is given as
$L=\frac{2QTS}{KQNT}=\frac{2S}{KN}$. Each identity submatrix is of size $g$, then the total number of $1$'s in each identity submatrix is $g$. There is no overlap between the identity submatrices, therefore the total number of $1$'s in computing matrix is $Sg$. Also, the number of ones in each column of computing matrix is $K-r$ as each subfile is stored in $r$ servers. Hence the total number of $1$'s in matrix $C$ is given by,
\begin{align}
\label{DC_eqn}
    Sg=N(K-r)
\end{align}

Using (\ref{DC_eqn}) we have, $L=\frac{2(K-r)N}{KgN}={\frac{2}{g}}{\left( 1- \frac{r}{K}\right)}.$

\end{IEEEproof}
We now give an example illustrating our new scheme, continuing from Example \ref{DC_matrix}, showing an identity submatrix cover for the computing matrix shown in that example. 

\begin{example}[Continuation of Example \ref{DC_matrix}]
The identity submatrices of the matrix in Example \ref{DC_matrix} shown using the $7$ shapes clearly form an identity submatrix cover consisting of non-overlapping identity submatrices. 

\begin{figure}[h]
  \centering
\includegraphics[height=3cm, width=5.8cm]{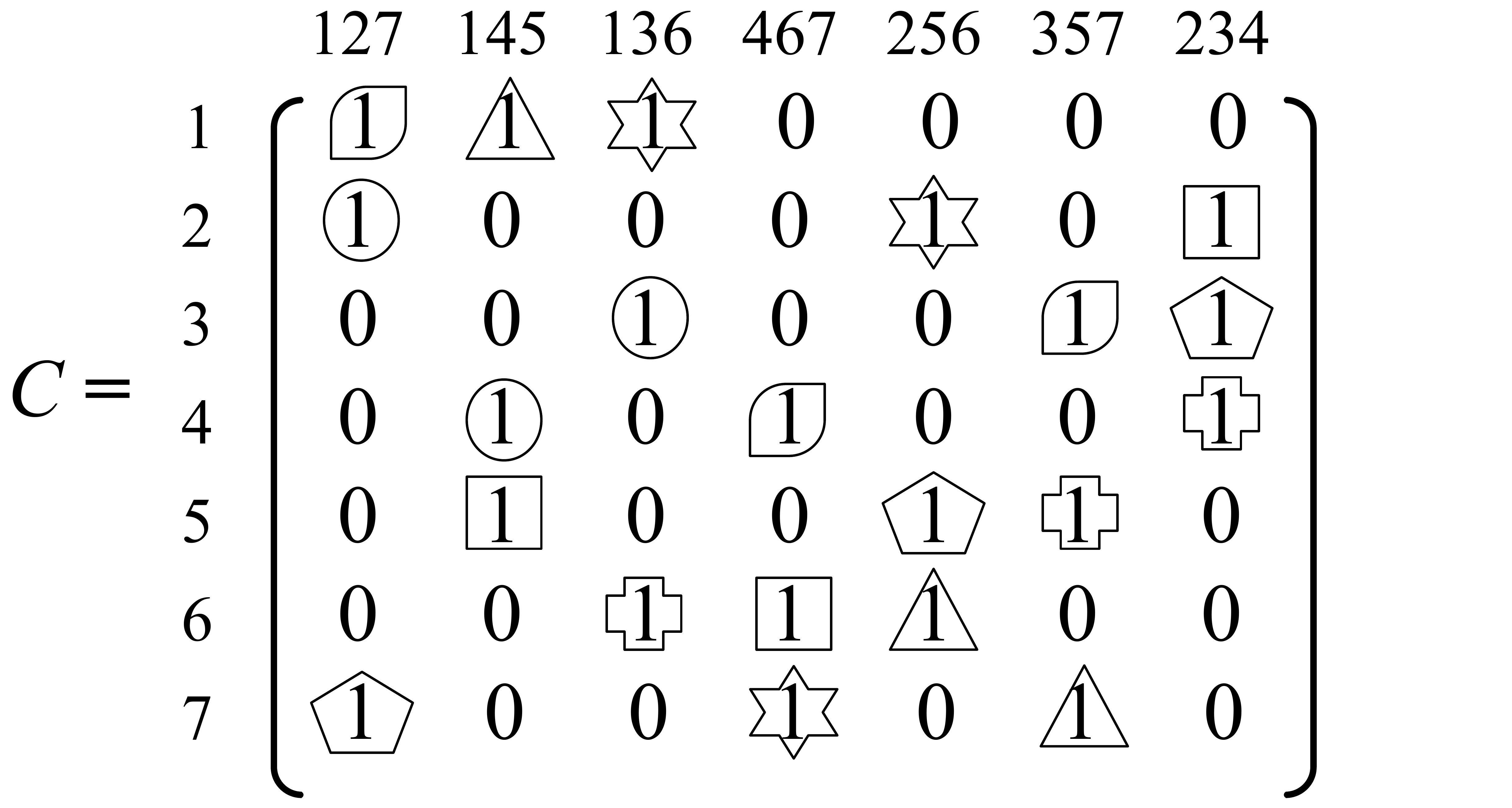}
\end{figure}
We now show one round of transmissions, consisting of two transmissions corresponding to one identity submatrix in the cover.  Let us consider one of the identity submatrix denoted as $C_2$, where $C_2$ is as below. \[
  \text{$C_2$} = \kbordermatrix{
    & 127 & 256  & 234 \\
    3 & 0 & 0 & 1 \\
    5 & 0 & 1 & 0 \\
    7 & 1 & 0 & 0 
  }
 \]
In $C_2$, one of the servers in $\{{3},{5},{7} \}$ will do the coded transmission and one will do the uncoded transmission. Let $Q=14$ (i.e., $\beta=2)$ then, ${\cal W}_{3} =\{3,10\}, ~{\cal W}_{5}=\{5,12\},~{\cal W}_{7}=\{7,14\}$.
IVAs needed at server ${3}$ to perform the reduce operation are $\{ v_{3,234}, ~v_{10,234} \}, 234\notin{\cal M}_{{3}}$. Similarly, IVAs needed at server ${5}$ and ${7}$ are $\{ v_{5,256}, ~v_{12,256} \}, 256\notin{\cal M}_{{5}}$ and $\{ v_{7,127}, ~v_{14,127} \},  127\notin{\cal M}_{{7}}$, respectively.
Let server $3$ does the coded transmission and server $5$ does the uncoded transmission. Then the coded transmission by server $3$ is $\{ v_{7,127} \oplus v_{5,256} ,~ v_{14,127} \oplus v_{12,256} \}$. Server $5$ already has $v_{7,127}$ and $v_{14,127}$ calculated in map phase, hence it can decode $v_{5,256}$ and $v_{12,256}$ from coded transmission. The IVAs missing at server $3$ can be received by uncoded transmission $\{ v_{3,234}, ~v_{10,234}  \}$ done by server $5$. 



In a similar way, transmissions are done corresponding to each such identity submatrix in identity submatrix cover. By Theorem \ref{DC_maintheorem}, the requisite IVAs are decoded.
\end{example}
The coded data shuffling scheme in \cite{RDC} achieves a communication load of $L=\frac{1}{r}(1-r/K)$, and this was shown to be optimal. We now recall this scheme and present the map phase using a binary computing matrix, and calculate the communication load achieved by our data shuffling scheme. This result is captured in the following corollary to Theorem \ref{DC_maintheorem}

\begin{corollary}
\label{corrMaNschemeOurShuffling}
For any positive integers $K$ and $r\in[K]$, there exists a $(K,\binom{K}{r},r)$-binary computing matrix, from which we get a distributed computing scheme on $K$ nodes with computation load $r$ and communication load $L=\frac{2}{r+1}\left(1-\frac{r}{K}\right)$, with file complexity $N=\binom{K}{r}$. Further, this load $L<2L^*(r)$, where $L^*(r)$ is the optimal rate for a given computation load $r$.
\end{corollary}
\begin{IEEEproof}
Let ${\cal K}=[K]$ denote the set of nodes, and ${\cal N}=\{f_A:A\in\binom{[K]}{r}\}$ denote the set of subfiles, where $\binom{[K]}{r}$ denotes the $r$-sized subsets of $[K].$ Consider the matrix $C$ of size $K\times \binom{K}{r}$ with $C(k,f_A)=0$ if $k\in A$, and $C(k,f_A)=1$ if $k\notin A.$ The matrix $C$ is thus a $(K,N=\binom{K}{r},r)$-binary computing matrix. A non-overlapping identity submatrix cover of this matrix can be easily obtained as follows. Consider a subset of size $r+1$ of $[K],$ denoted by $B$. It is easy to check that the collection of rows defined by $B$ and the columns $B\backslash\{k\}:k\in B$ define an identity submatrix, which we denote by $C_B$. Further for each such $B$, the submatrix $C_B$ is of size $r+1$, and it is straightforward to check that these matrices are non-overlapping. Further, for $C(k,f_A)=1$ entry is covered by precisely that identity submatrix defined by $\{k\}\cup A.$ Thus, the collection of identity submatrices $\{C_B : B\in \binom{[K]}{r+1}\}$ is a non-overlapping identity submatrix cover of $C$, with $g=r+1$. Thus, by Theorem \ref{DC_maintheorem}, our data shuffling scheme on this binary computing matrix $C$ achieves a communication load $L=\frac{2}{r+1}(1-\frac{r}{K}).$ As  $L^*(r)=\frac{1}{r}(1-r/K)$ is known from \cite{RDC} to be the optimal communication load for computation load $r$, by comparing the two expressions we see that $L=\frac{2r}{r+1}L^*(r)<2L^*(r).$
\end{IEEEproof}
\subsubsection{Advantages of our shuffling scheme over the optimal scheme in \cite{RDC}}
\label{advantagessubsubsection}
Corollary \ref{corrMaNschemeOurShuffling} shows that our scheme has a higher communication load than the optimal scheme in \cite{RDC}. We now discuss some advantages of our scheme over the shuffling scheme in \cite{RDC}. During the data shuffling phase of the optimal-load scheme in \cite{RDC}, the IVAs have to be further subdivided into $r$ smaller chunks of length $\frac{T}{r}$ bits each. Then, for each subset $B$ of $[K]$ of size $r+1$, every server in $B$ encodes a set of $r$ chunks and broadcasts it to the other servers in $B$. The further chunking of the IVAs is absent in our scheme. This further dividing of the IVAs into $r$ smaller chunks incurs multiple costs including in coordination, indexing, switching, etc. which we now discuss. As a result of avoiding this IVA chunking, we refer to our scheme as a low complexity scheme compared to those in \cite{RDC}. 
\begin{itemize}
    \item Firstly, to identify the chunks, some indexing is required. This chunk-indexing cost is additional over and above the original file-complexity $N$. Our new scheme avoids this further chunking, and hence does not incur this cost. 
    \item Then, it is a requirement that multiple servers which have computed the same IVA in the map phase employ the exact IVA file chunking. If this is not done, decoding will not be possible. This decentralized IVA-chunking is thus unlike the original file complexity $N$, which is done prior to the placement in the storage of the nodes for mapping, and possibly in a single machine. This decentralized IVA-chunking of \cite{RDC} therefore requires some further coordination to establish agreement amongst the various nodes compared to our scheme.  
    \item Further, reading a large number of smaller sized chunks from the actual memory device (for instance, a hard disk or a flash drive) is more time and power consuming when compared to obtaining a smaller number of larger sized reads (our reads would be entire IVAs, i.e., $r$-times the size of the read in the scheme of \cite{RDC}). 
    \item Finally, suppose the transmissions at any node happen in a sequential manner following their occurrence in different sets $B\subseteq [K]$ of size $r+1$. Then since every node in $B$ participates in the transmission corresponding to $B$, this incurs the additional cost of turning the transmitting device ON and OFF a large number ($r\binom{K}{r}$) of times. However, in our scheme, only $2$ servers participate in the transmission round corresponding to any $B$. This means that we incur a cost of $2\binom{K}{r}$ number of switchings only. 
\end{itemize}
\subsection{Communication Load balancing of Scheme in Theorem \ref{DC_maintheorem}}
\label{loadbalancingsubsec}
The data shuffling scheme according to Theorem \ref{DC_maintheorem} ensures that only $2$ servers have to transmit for each identity submatrix in the identity submatrix cover $\mathfrak{C}$. As Lemma \ref{lemma_identitysubmatrix} chooses them arbitrarily, this may lead to a situation of imbalance in the communication load, i.e., some servers could be transmitting more bits while others transmit much less, or even don't transmit at all. This imbalance of network traffic may lead to other problems like node failures due to excess load, overall performance loss, etc. The following result shows that this problem of load imbalance can be rectified (provided some simple condition holds) by identifying two perfect matchings on an appropriately defined bipartite graph, which can be done in polynomial time in the parameter $S$. These conditions hold for many constructions we present, as well as those in literature such as that in \cite{RDC}.
\begin{theorem}
\label{theorem_load balancing}
Let $C$ be a $(K,N,r)$-binary computing matrix with an non-overlapping identity cover $\mathfrak{C}=\{C_1,\hdots,C_S\},$ such that the size of each identity submatrix $C_i$ is $g\geq 2$ representing a computing system to reduce $Q$ functions. Then the coded data shuffling scheme in Theorem \ref{DC_maintheorem} is achievable with the property of load balancing, i.e. the total number of bits transmitted by each node is exactly $\frac{2S\beta T}{K}$ (where $\beta=Q/K$) out of which $\frac{S\beta T}{K}$ bits correspond to coded bits and the other $\frac{S\beta T}{K}$ bits correspond to uncoded bits, if (a) $\gamma\triangleq \frac{S}{K}$ is an integer, and (b) if each server $k$ appears in the row indices of the same number of identity submatrices in $\mathfrak{C}.$

\end{theorem}
\begin{IEEEproof} 
We first construct a bipartite graph ${\cal B}$. The set of left vertices of $\cal B$ are the set of servers (${\cal K}$) repeated $\gamma$ times, and is denoted as $\{k^{j}: k \in {\cal K}, j\in [\gamma]\}$. The right vertices are the indices of the identity submatrices in $\mathfrak{C}$ respectively. The edges are defined as follows. An edge between $k^j $ and $ C_{i}$ exists if and only if server $k$ is present in the row of identity submatrix $C_i$. Since the size of any identity submatrix in $\mathfrak{C}$ is $g$, the graph $\cal B$ is thus right regular with degree $g\gamma$, which means that the graph is biregular with degree $g\gamma$ as the cardinality of left vertices and right vertices are the same (namely, $\gamma K=S$), and by property (b).

A perfect matching on a bipartite graph is a matching $M$ (a collection of edges with no common vertices) such that every vertex in the graph is incident on at least one edge in the matching $M$. For regular bipartite graphs with $n$ vertices, a perfect matching can be found in time $O(n\log n)$ \cite{perfectM}. If such a perfect matching is found in $\cal B$, then for each $k\in{\cal K}$, each left vertex in the set $\{k^j:j\in[\gamma]\}$  is matched with precisely one right vertex, and thus the vertices $\{k^j:j\in[\gamma]\}$ are matched to right vertices (i.e., $\gamma$ identity submatrices), say $\{C_{k_i}:i\in[\gamma]\}$. In that case, we make the vertex $k$ responsible for the coded transmission corresponding to the identity submatrices $C_{k_i}:i\in[\gamma]$. Therefore each vertex $k$ is responsible for $\gamma$ coded transmissions. 

Now to define the server identity submatrix pairing for uncoded transmissions, we first obtain a new bipartite regular graph of degree $(g\gamma-\gamma)$ from $\cal B$ by removing some edges corresponding to the already-found perfect matching. Let for some $k,p$ and $i$, there exists an edge between $k^{p}$ and $C_{i}$ in perfect matching we have found, then we will remove all the edges between $k^{j}$ and $C_{i},~ \forall j\in [\gamma]$ from ${\cal B}$ . It is easy to see that the degree of each vertex of ${\cal B}$ is reduced to $g\gamma - \gamma=\gamma(g-1)$.  Note that as $g\geq 2$, we must have $\gamma(g-1)\geq 1$. Thus we have a new graph which is regular with degree $\gamma(g-1)$. Hence, we can find a perfect matching on this graph once again, using the algorithm in \cite{perfectM} for instance. By a similar argument as in the previous paragraph, for each $k\in{\cal K}$, we can get another set of $\gamma$ identities associated to it arising out of the new perfect matching say $\{C_{k_j}:j\in[\gamma]\}$. However, since the edges as described above are removed, we must have that $\{C_{k_i}:i\in[\gamma]\}\cap\{C_{k_j}:j\in[\gamma]\}=\phi$, i.e., none of the identities associated to $k$ in the first matching are repeated in the second. As a result, each $k\in{\cal K}$ can be assigned $\gamma$ uncoded transmissions corresponding to $\gamma$ identity submatrices which are all distinct from the $\gamma$ identities that $k$ has already been assigned to do coded transmissions for. 
Now, for each identity submatrix (right vertex), we have thus got two edges arising from the two perfect matchings obtained in the above manner, and these two must necessarily be incident on two nodes $k_1^{j_1},k_2^{j_2},$ where $k_1\neq k_2$. Thus, we have identified two distinct servers $k_1$ and $k_2$ from the first and second perfect matching doing the coded and uncoded transmission for each identity submatrix respectively. Combining the identification of these server nodes which are responsible for the coded and uncoded transmissions with the arguments of Theorem \ref{DC_maintheorem} which continue to hold as is, the proof is complete.

\end{IEEEproof}

\begin{figure}
  \centering
    \begin{subfigure}[b]{0.2\textwidth}
                \centering
                \includegraphics[height=4cm, width=3cm]{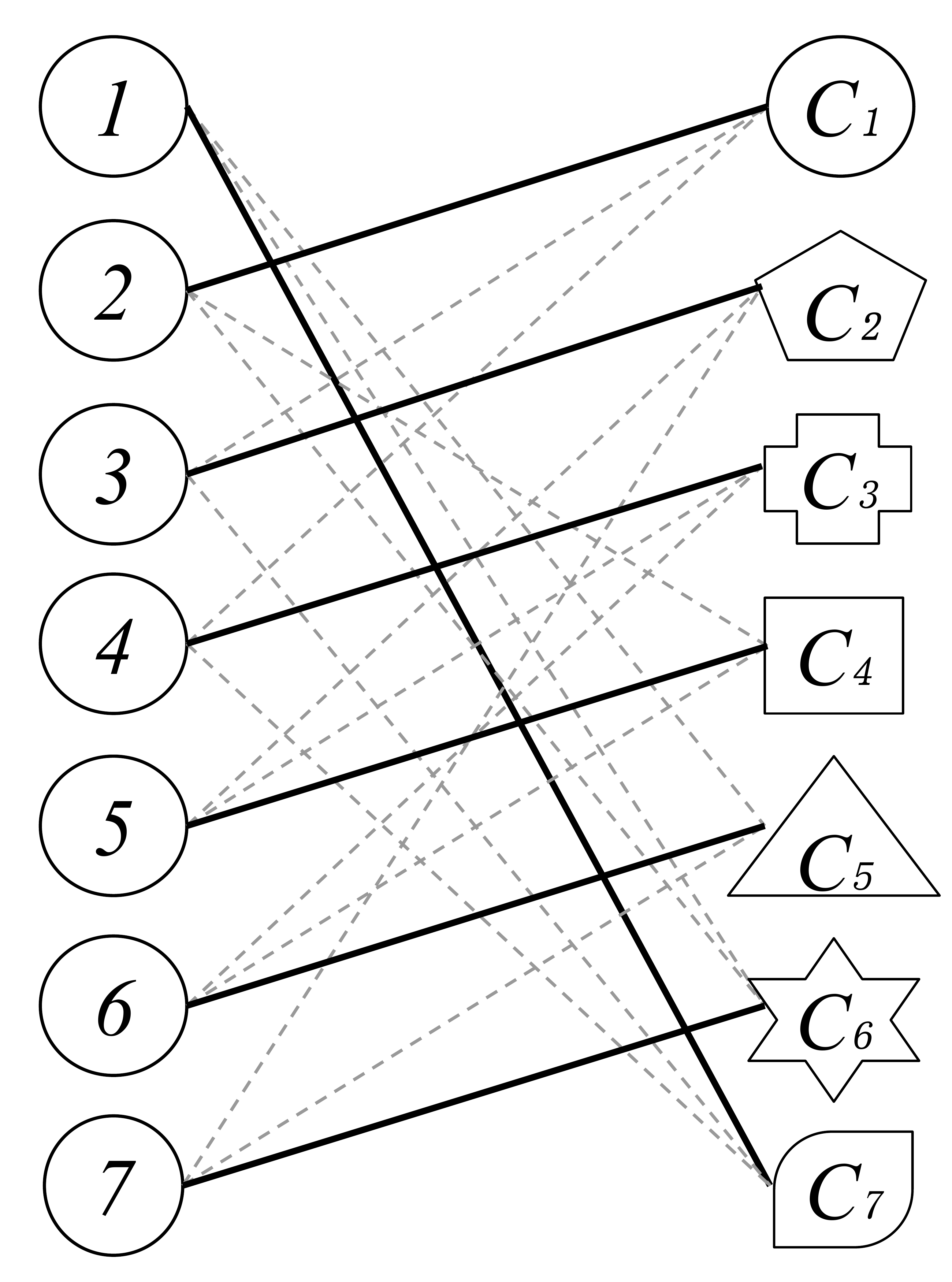}
        \end{subfigure}
        \begin{subfigure}[b]{0.19\textwidth}
                \centering
                \includegraphics[height=4cm, width=3cm]{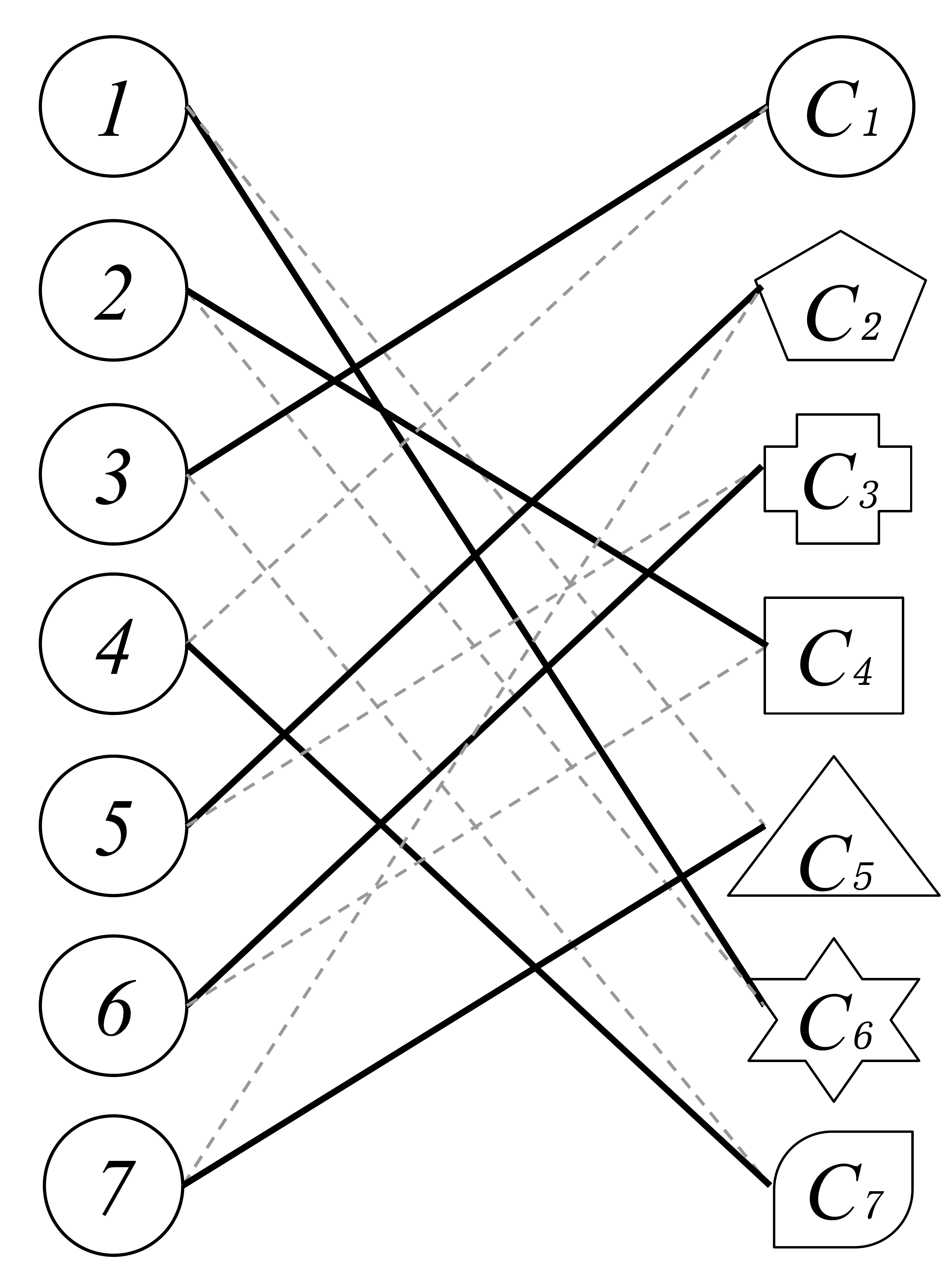}
        \end{subfigure}%
        \caption{Bipartite graphs based on the binary computing matrix in Example \ref{DC_matrix} illustrating Theorem \ref{theorem_load balancing}}
        \label{DC_graph}
       \end{figure}

\begin{example}
Let us consider the identity submatrix given in Example \ref{DC_matrix} for a distributed computing scenario where $K=7$ and $N=7$, for which an identity submatrix cover with $S=7$ matrices is shown in Example \ref{DC_matrix}. Thus $\gamma=\frac{S}{K}=1.$ For simplicity, we assume that $K=Q$ functions need to be reduced, thus $\beta=1$. The first figure on the left in Fig. \ref{DC_graph} shows the bipartite graph as constructed in Theorem \ref{theorem_load balancing}, with the users vertices on the left and the 7 identities on the right shown using the shapes. The bold edges in the first figure denote the first perfect matching obtained, using which coded transmissions are assigned. For instance, the user $6$ participates in the coded transmission with respect to the identity  submatrix corresponding to the shape `triangle' in Example \ref{DC_matrix}. After deleting these edges, we get the bipartite graph on the right, which is again a regular graph. The bold edges on this graph denote the perfect matching corresponding to the uncoded transmissions. Thus, each user is seen to participate in $2$ transmissions in this case, as $\beta=1,$ and $S=K=7$, transmitting $2T$ bits in total as given in Theorem \ref{theorem_load balancing}.

\end{example}

\subsection{Low file complexity ($N$) schemes based on binary matrices from combinatorial designs}
\label{combinatorialdesigns}
In \cite{SSP} (and its extended version \cite{SSPArXiv}), binary matrices with  constant row weight were derived using a variety of combinatorial designs including Balanced Incomplete Block Designs (BIBDs), symmetric BIBDs, $t$-designs, and transversal designs. 
 For instance, the definition of a BIBD is as follows. 
 
\begin{definition}[BIBD \cite{Drs}]
Let $v, k$ be positive integers such that $v>k \geq 2$. A $(v, k,1)$-BIBD is a design $(\cal X, A)$ such that the following properties are satisfied: \newline (1). ${|\cal X|} = v$,
 \newline   (2). Each block contains exactly $k$ points, and
  \newline   (3). Every pair distinct points is contained in exactly one block.
  \end{definition}

It turns out that the number of blocks in $(v,k,1)$-BIBD is $\frac{v(v-1)}{k(k-1)}$ (please see \cite{Drs} for more details). Using the incidence matrix of the BIBD, we obtain a binary matrix of size $v\times \frac{v(v-1)}{k(k-1)}$, which has constant column weight of $k$, and hence $r=v-k$. Thus, this matrix gives us a $(K=v,N=\frac{v(v-1)}{k(k-1)},r=v-k)$-computing matrix. This is tabulated in the first row of Table \ref{DC_table} as Scheme I. An identity submatrix cover of this matrix is shown in Section V of \cite{SSPArXiv}, consisting of matrices of size $g\triangleq \frac{v-1}{k-1}$. Plugging this in Theorem \ref{DC_maintheorem} gives us the communication load as $\frac{2k(k-1)}{v(v-1)}$. This is captured in the sixth column of Table \ref{DC_table}. The optimal scheme for the parameters $K=v$ and $r=(v-k)$ is however the scheme of \cite{RDC}, which achieves a load $\frac{1}{r}(1-r/K)=\frac{1}{v-k}\left(1-\frac{v-k}{v}\right)=\frac{k}{(v-k)v}$. This is clearly smaller than the load of our BIBD based scheme. However, this is achieved with a high file complexity of $\binom{K}{r}=\binom{v}{k}$, and with further IVA chunking. Our BIBD-based computing scheme however has a file complexity $N=\frac{v(v-1)}{k(k-1)},$ which is much smaller than $\binom{v}{k}$ in general, and our data shuffling scheme avoids the IVA chunking as well. 

In the same manner, using the various binary matrices derived from symmetric BIBDs (Section VI in \cite{SSPArXiv}), $t$-designs (Section VII in \cite{SSPArXiv}), and transversal designs (Section VIII in \cite{SSPArXiv}), we can obtain binary matrices which have constant row weight. By some simple arguments we can show that these matrices have constant column weight as well, which leads us to consider them as binary computing matrices. Using these matrices, and using our low-complexity data-shuffling schemes, the parameters obtained for the computing schemes are listed in the first six columns of Table \ref{DC_table}. In each case, we see that the file complexity is quite small compared to the file complexity for the same $K,r$ values in the optimal load scheme of \cite{RDC}, however it can be seen that we pay a price in the load achieved. 
We have skipped the definitions, specific calculations and properties, and the proofs of some claims above, with respect to the various other designs mentioned here, as they are more cumbersome than illuminating. Most of these however are verbatim from \cite{SSPArXiv} or a simple inference, and hence we refer the reader to \cite{SSPArXiv} for these.

\begin{table*}
 \tabulinesep=1.0mm
\centering
\captionof{table}{Parameters of Distributed Computing Scheme based on Combinatorial Designs constructed in \cite{SSP}. Note that the parameters $v,k,t,n$ correspond to those of the respective combinatorial designs}
\begin{tabular}{|c|c|c|c|c|c|c|}
 \hline
 
 & \textbf{Combinatorial} & \textbf{Number of}   & \textbf{Number of } & \textbf{Computation Load} & \textbf{Communication Load}   &  \textbf{Communication Load}     \\

S.No. & \textbf{ Designs} & \textbf{servers} & \textbf{subfiles}  & $\textit{r}$ &  \textbf{for non straggler case} &  \textbf{for $K-\kappa$}   \\
 
 & & $\textit{K}$ & $\textit{N}$ & & & \textbf{full straggler case}\\
 \hline

\RN{1} &\textit{ BIBD (${\lambda=1}$)}&$v$&$\frac{v(v-1)}{k(k-1)}$&$(v-k)$&$\frac{2k(k-1)}{v(v-1)}$& $\frac{2k(k-1)}{\kappa (v-1)}$ \\
\hline

\RN{2}&\textit{Symmetric BIBD (${\lambda=2}$)}&$v$&$kv$&$(v-k+1)$&$\frac{2}{v}$ & $\frac{2}{\kappa}$ \\
\hline

\RN{3}& ${t}$-\textit{design (${\lambda=1}$) (Scheme $1$)} &$\binom{v}{t-1}$&$\frac{{\binom{v}{t}k}}{\binom{k}{t}}$&$\binom{v}{t-1}-\binom{k-1}{t-1}$& $\frac{2(v-t+1)\binom{k-1}{t-1}^2}{v\binom{v-1}{t-1}^2} $ & $\frac{2\binom{k-1}{t-1}^2}{\kappa \binom{v-1}{t-1}} $ \\
\hline

\RN{4}& ${t}$-\textit{design (${\lambda=1}$) (Scheme $2$)} &$v$&$\binom{v}{t}$&$(v-t)$& $\frac{2t}{v(v-t+1)}$ & $ \frac{2t}{\kappa (v-t+1)}$\\
 \hline

\RN{5}& \textit{ Transversal Design (${\lambda=1}$)} &  $n^{2}$ & $kn$ &$n(n-1)$ & $\frac{2}{n^2}$ &  $\frac{2}{\kappa}$\\
\hline
\end{tabular}
\label{DC_table}
\end{table*}

\section{Extension to Straggler Scenario}
\label{fullstragglersec}
One of the practical challenges in the distributed computing framework is the presence of \textit{straggling nodes}. Straggling nodes are the nodes that perform operations slower than the other nodes. In this section, we utilize the advantage of our scheme (Theorem \ref{DC_maintheorem}, Lemma \ref{lemma_identitysubmatrix}) that only two servers are involved to communicate in each transmission round (corresponding to one identity submatrix). This advantage is used for straggler robustness upto a fixed number of stragglers, namely $g-2$, where $g$ is the size of any identity. In the \textit{full straggler} scenario, which is our main contribution in this section, the straggling nodes are slow to the extent they will not be able to complete any map task assigned to them, i.e. that can be considered as failed nodes. Thus, they are not involved in the map, shuffle or the reduce phase. This setting was assumed in \cite{QYS}, the scheme from which we shall compare with. We also make some comments on the partial straggler scenario at the end of the paper. 

In the straggler scenario, our goal for the shuffling phase remains the same: to exchange messages between the nodes so that the IVAs for the reduce phase at the respective nodes are available. We thus redefine the communication load for the straggler scenario as 
\begin{align*}
L(\kappa)= \small \frac{\text{Number of bits transmitted in shuffle phase in `worst case'}}{\text{$QNT$}}
\end{align*}
where the `worst case' refers to the worst subset of $K-\kappa$ stragglers (that subset which creates the largest load). 



\label{DC_Full Stragglers}
In this section, we will discuss how to use the binary computing matrix to deal with full stragglers. Our scheme is robust upto $g-2$ stragglers. We consider a setup that as soon as $\kappa, \kappa\in{\cal K}$ servers complete the map operation, data shuffling phase can start.  We call this set of $\kappa $ servers to be \textit{surviving servers} while the rest $K-\kappa \in [0:g-2]$ servers are the full stragglers. For simplicity, we assume that $\frac{Q}{\kappa}$ is an integer, and thus $Q$ functions are evenly distributed among the $\kappa$ surviving servers. Below we describe a scheme which is robust for $g-2$ full stragglers. 
\begin{theorem}
\label{fullstragglertheorem}
Consider a \textit{computing matrix} of size $K\times N$ with a non-overlapping identity submatrix cover $\mathfrak{C}= \{C_1,C_2,..,C_S\}$ where the size of each identity submatrix is $g,\textcolor{red}{g\geq 2}$. Then, there exists a distributed computing scheme with $K$ nodes that is robust for $K-\kappa \in [0:g-2]$ full stragglers, attaining computation load $=r$ and communication load $L(\kappa)={\dfrac{2}{g}}{\left( \dfrac{K}{\kappa}- \dfrac{r}{\kappa}\right)},$ with file complexity $N$. 
\end{theorem}
\begin{IEEEproof}
Suppose some arbitrary $K-\kappa$ set of nodes failed. Since $K-\kappa\leq g-2,$ at least $2$ servers must still survive with respect to the rows of any identity submatrix $C_i$ in the cover $\mathfrak{C}.$ Thus, two servers for transmissions as in Lemma \ref{lemma_identitysubmatrix} are available for each identity submatrix. Hence a similar scheme as in Theorem \ref{DC_maintheorem} will be feasible, with the difference that since each server is allocated to reduce $\frac{Q}{\kappa}$ functions rather than $\frac{Q}{K}$ as in Theorem \ref{DC_maintheorem}. Following the similar arguments as Theorem \ref{DC_maintheorem}, the communication load in this case is calculated as follows.


\begin{center}
$\text{}~L =\dfrac{2QST}{\kappa QNT}=\dfrac{2S}{\kappa N}$ 
\end{center}
Using (\ref{DC_eqn}) we have,
\begin{center}
$L=\dfrac{2(K-r)N}{\kappa gN}={\dfrac{2}{g}}{\left( \dfrac{K}{\kappa}- \dfrac{r}{\kappa}\right)} \cdot$
\end{center}
Noting that we considered an arbitrary set of stragglers completes the proof.
\end{IEEEproof}
\begin{table}[ht]

\tabulinesep=1.0mm
\centering
\captionof{table}{Communication Load comparison for Distributed Computing Scheme with full stragglers. }
\begin{tabular}{ |M{2.0cm} |M{1.4cm}| M{1.8cm}|M {1.8cm}|  }

 \hline
 
\textbf{Distributed computing Parameters of MAN-PDA } & \textbf{Number of non stragglers} $\kappa$ & \textbf{Optimal Communication Load in \cite{QYS}}  & \textbf{Communication Load in Theorem \ref{fullstragglertheorem}}\\
 \hline

$K=5,  r=2 $, $N=10, $  $g=3$ & $ 4$ & $0.45$ & $0.5$\\
\hline

$K=7,  r=4 $, $N=35,$  $g=5$ & $ 5$ & $0.169$ & $0.24$ \\
\hline

$K=7,  r=4 $, $N=35,$  $g=5$ & $ 4$ & $0.2428$ & $0.3$ \\
\hline

$K=10,  r=3 $, $N=120,$ $g=4$ & $ 8$ & $0.3305$ & $0.4375$ \\
\hline

\end{tabular}
\label{DC_table2}
\end{table}

In \cite{QYS} (and the extended version \cite{QYSArXiv}), the authors use the scheme of \cite{RDC} (as given in Corollary \ref{corrMaNschemeOurShuffling}) to achieve robustness against any set of $K-\kappa$ stragglers, where $K-\kappa \leq r-1$, and achieve a load given by the expression 
$$L^*(\kappa)=\left(1-\frac{r}{K}\right) \sum\limits_{\substack{i=r+\kappa-K}}^{min\{r,\kappa-1\}} {\frac{1}{i}\dfrac{\binom{r}{i}\binom{K-r-1}{\kappa-i-1}}{\binom{K-1}{\kappa-1}}}.$$ In fact, this load happens to be optimal given parameters $K,r$, and number of stragglers $K-\kappa\leq r-1$ \cite{QYSArXiv}.
Following Corollary \ref{corrMaNschemeOurShuffling} and using Theorem \ref{fullstragglertheorem}, we see that the same scheme of \cite{RDC} which corresponds to $(K,N=\frac{K}{r},r)$-binary computing matrix (with $g=r+1$), we can obtain a distributed computing scheme that is robust against $K-\kappa$ stragglers (where $K-\kappa\leq g-2=r-1$), and has a load $L=\frac{2}{\kappa(r+1)}(K-r)$. 

We plug different values of $K,r,\kappa$ in the scheme of \cite{QYS} (which is shown to be optimal) and also for our scheme and compare communication loads for full straggler case in Table \ref{DC_table2}. We note that the load of our scheme is higher, though still comparable. Note that the file complexity is $N=\binom{K}{r}$ for both our scheme and that in \cite{QYS}. However, our scheme retains the advantages as mentioned in Section \ref{advantagessubsubsection}. We further tabulate in Table \ref{DC_table} (last column) the load achieved by our low file-complexity computing schemes from the combinatorial designs of \cite{SSP}, which are obtained based on Theorem \ref{fullstragglertheorem} and the load for the non-straggler case in the sixth column (The $(K-\kappa)$ straggler load is $\frac{K}{\kappa}$ times the non-straggler load, as seen from Theorem \ref{fullstragglertheorem} and Theorem \ref{DC_maintheorem}). 

\begin{remark}
Finally, we remark on the partial straggler scenario. Our scheme based on Theorem \ref{DC_maintheorem} can be extended to the case of partial stragglers as well, where we model the partial stragglers as being capable of mapping only a subset of the subfiles in the map phase, cannot execute the shuffle phase, but are still responsible for the reduce phase. In that case, in the map phase, the partial stragglers (upto $g-2$ in number) are required to map only a corresponding subset of the subfiles that they have in their storage with respect to decoding their required IVAs and nothing else. The communication load in this case remains the same as the no-straggler load.
\end{remark}
\section{Conclusion}
\label{conclusion}
In this work, we have presented a binary matrix model for distributed computing, and a simple new coded data shuffling scheme for the same. The presented scheme achieves a load that is strictly less than twice that of the optimal scheme, for the same map phase designed by the optimal scheme. However, our scheme has advantages in the complexity of the data-shuffling scheme, since it avoids the chunking of the IVAs into smaller pieces. Under some mild conditions, we achieve load balancing of our schemes amongst the servers. Further, we also present distributed computing schemes using binary matrices arising from some combinatorial designs which have the advantage of very small file-complexity schemes when compared to the optimal scheme for similar values of $K,r$, at the cost of having a higher rate. We extend our scheme to the full straggler scenario as well, the numerical comparisons for which show marginal increase in the communication load compared to the optimal case. It would be interesting to further probe the use of binary matrices and combinatorial designs for the distributed computing scenario for constructing schemes which have advantages both in the file complexity as well as in the computation-communication loads. 

\bibliography{root.bib}
\bibliographystyle{ieeetr}
\end{document}